\documentclass[twocolumn,showpacs,prl,english,showkeys,aps,unsortedaddress,superscriptaddress,nofootinbib]{revtex4-2}
\usepackage{amsmath}
\usepackage{amssymb}
\usepackage{epsfig}
\usepackage{graphicx}
\usepackage[colorlinks]{hyperref}
\usepackage[T1]{fontenc}
\usepackage[utf8]{inputenc}
\usepackage{float}
\usepackage{color}
\usepackage[normalem]{ulem}
\usepackage{tikz}
\usetikzlibrary{positioning}

\usepackage{comment}
\usepackage[dvipsnames, svgnames]{xcolor}
\usepackage[compat=1.1.0]{tikz-feynman}
\usepackage{lipsum}

\begin{document}

\title{Fate of the Ising universality class under nonreciprocal interactions}

\author{M. Akritidis$^*$}
\affiliation{Institut de Math\'{e}matiques de Bourgogne, Universit\'{e} de Bourgogne Europe, 21078 Dijon, France}

\author{A. Garcés$^*$} 

\affiliation{Computing and Understanding Collective Action (CUCA) Lab, Condensed Matter Physics Department, Universitat de Barcelona, Martí i Franquès 1,
E08028 Barcelona, Spain}

\affiliation{University of Barcelona Institute of Complex Systems (UBICS), Martí i Franquès 1, E08028 Barcelona, Spain}

\author{A. Vasilopoulos$^*$}
\affiliation{School of Mathematics, Statistics and Actuarial Science, University of Essex, Colchester CO4 3SQ, United Kingdom}

\author{M. Carosi$^*$}
\affiliation{International Centre for Theoretical Physics Asia-Pacific (ICTP-AP), University of Chinese Academy of Sciences, 100190 Beijing, China}

\author{D. Levis}\email{levis@ub.edu}

\affiliation{Computing and Understanding Collective Action (CUCA) Lab, Condensed Matter Physics Department, Universitat de Barcelona, Martí i Franquès 1,
E08028 Barcelona, Spain}

\affiliation{University of Barcelona Institute of Complex Systems (UBICS), Martí i Franquès 1, E08028 Barcelona, Spain}

\author{N.~G. Fytas}\email{nikolaos.fytas@essex.ac.uk}
\affiliation{School of Mathematics, Statistics and Actuarial Science, University of Essex, Colchester CO4 3SQ, United Kingdom}

\def\thefootnote{*}
 
\date{\today}

\begin{abstract}
We study the critical behavior of a two-dimensional Ising model with nonreciprocal vision-cone interactions, which explicitly violate reciprocity and detailed balance. Extensive Monte Carlo simulations and dynamic renormalization-group analysis show that the asymptotic critical exponents remain fully consistent with the equilibrium Ising universality class over a broad range of nonreciprocal coupling strengths $\lambda$. In contrast, dimensionless quantities such as the Binder cumulant and the correlation-length ratio display pronounced anisotropic nonequilibrium corrections and systematically deviate from their equilibrium Ising values. The renormalization-group flow further demonstrates that the nonreciprocal perturbation is irrelevant at the Wilson--Fisher fixed point while generating a finite shift of the critical temperature proportional to $\lambda^2$. Our results demonstrate the remarkable robustness of two-dimensional Ising criticality against this class of directional interactions.
\end{abstract}

\maketitle

\footnotetext{These authors contributed equally to this work.}

\emph{Introduction.---} Reciprocity is a central ingredient of equilibrium statistical mechanics, underlying the existence of effective Hamiltonians, detailed balance, and conventional thermodynamic descriptions. In recent years, however, there has been growing interest in systems with explicitly non-reciprocal (NR) interactions, ranging from active and driven matter to biological collectives and non-Hermitian condensed-matter systems~\cite{dadhichi2020nonmutual, okuma2023non, Loos23, pisegna2024emergent,  Rouzaire25,Garces2025,Avni25,popli2025ordering, Bandini25, rouzaire2026dynamics,carosi2026time, fruchart2026nonreciprocal}. In such systems, the interaction exerted by one degree of freedom onto another is not necessarily matched by the reverse interaction, leading to genuinely non-equilibrium steady states characterized by probability currents in configuration space and unconventional collective behavior~\cite{SCHMITTMANN95,Zia02,tauber_book,Klapp23,Ma25}.

A fundamental open question concerns the robustness of equilibrium universality classes under NR dynamics. Equilibrium critical behavior is largely determined by dimensionality, symmetries, and conservation laws~\cite{Wilson74,Hohenberg77,justin_book,amit_book}, yet NR interactions may dynamically generate directional information flow, effective anisotropies and violations of detailed balance capable of modifying the large scale critical properties. Early field-theoretical studies~\cite{Grinstein85} argued that local non-equilibrium perturbations do not alter the critical behavior of  a scalar order parameter field theory under non-conserved dynamics, provided the underlying $\mathbb{Z}_2$ up-down symmetry is maintained. 
This conclusion was later extended near the upper critical dimension $d_{\rm c} = 4$, showing that the kinetic Ising universality class remains stable against a wider class of non-equilibrium perturbations eventually breaking the $\mathbb{Z}_2$ symmetry~\cite{Bassler94}. Despite these results, explicit microscopic realizations of NR lattice models displaying controlled nonequilibrium criticality remain scarce.

In this work, we investigate a NR extension of the two-dimensional ($2d$) Ising model with vision-cone interactions~\cite{Garces2025}. Each spin interacts asymmetrically with its nearest neighbors according to its local direction (within a given \emph{vision-cone}), explicitly breaking reciprocity and preventing the existence of a global Hamiltonian. At the coarse-grained level, the model generates anisotropic non-equilibrium  nonlinearities of the type previously discussed in field-theoretical studies~\cite{Bassler94, carosi2026time}, while preserving a continuous second-order phase transition. Using large-scale Monte Carlo (MC) simulations together with finite-size scaling analysis and Renormalization-Group (RG) arguments, we show that the asymptotic critical behavior remains governed by the $2d$ Ising universality class over a broad range of NR interaction strengths. Critical exponents remain fully consistent with their exact Ising values despite the explicit violation of detailed balance and reciprocity. At the same time, dimensionless quantities such as the Binder cumulant and the correlation-length ratio display pronounced and systematic deviations from their equilibrium Ising values.

\begin{figure}[ht!]
    \centering
    \includegraphics[width=1.0\linewidth]{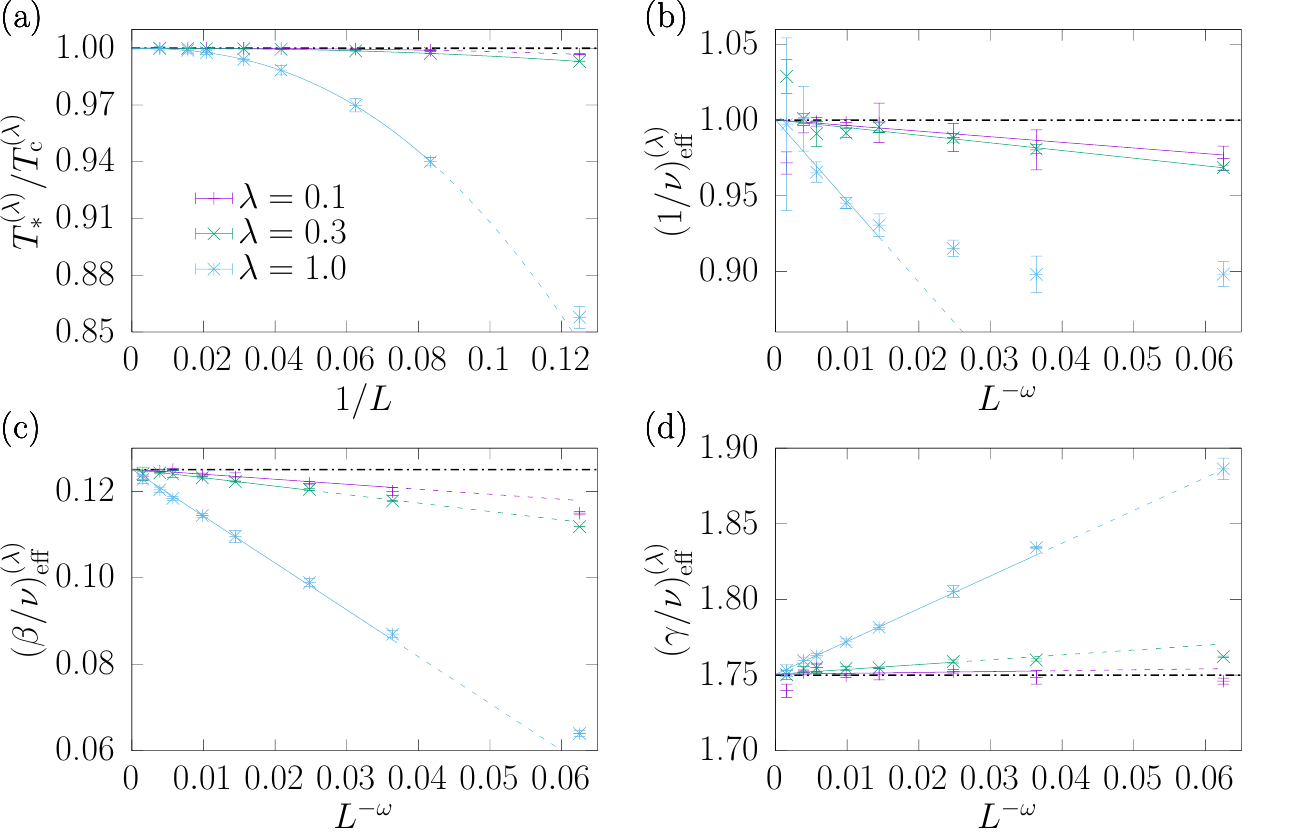}
     \caption{Finite-size scaling analysis of (a) the crossing temperatures $T_{\ast}$ of the Binder cumulant $U_4$ for system sizes $(L,2L)$, together with effective estimates of (b) $1/\nu$ obtained from ratios of  $dU_4/dT$ at the crossings, (c) $\beta/\nu$ from the magnetization $m$, and (d) $\gamma/\nu$ from the  susceptibility $\chi$. The vertical axis in panel (a) has been normalized by the corresponding extrapolated value of $T_{\rm c}$. In panels (b)--(d), the horizontal dot-dashed lines indicate the exact values of the $2d$ Ising model. Panels (b)--(d) are shown together with simultaneous joint fits to the finite-size scaling forms discussed in the text.}
    \label{fig:MC:Tc_exponents}
\end{figure}

Our results indicate that the NR vision-cone interactions act primarily through anisotropic scaling corrections that strongly affect finite-size amplitude ratios, while leaving the underlying Ising fixed point stable, the location of which is shifted toward higher temperatures as non-reciprocity increases. More generally, the present work provides direct microscopic evidence that equilibrium Ising criticality remains robust even in the presence of intrinsically NR dynamics, helping to clarify the role of nonequilibrium perturbations in critical many-body systems. 

\emph{Lattice model and MC simulations.---} We study the NR extension of the nearest-neighbor Ising model on the square lattice with linear dimension $L$ and visual-cone interactions introduced in ~\cite{Garces2025}. The dynamics is defined through the local effective energy field
\begin{equation}
E_i=-\sigma_i\sum_{j=1}^{N}J_{ij}\sigma_j,
\label{eq:MC:local_energy}
\end{equation}
where $\sigma_i=\pm 1$ denotes the spin at site $i$, $N = L^{2} $ the total number of spins, and $J_{ij}$ is the interaction strength between nearest-neighbors sites $i$ and $j$. Defining the unit vectors $\hat e_{ij}  = \pm\hat x,\pm\hat y$ 
connecting two nearest-neighbor sites and the polarization vector $\hat p_i=\sigma_i(\hat x+\hat y) / \sqrt{2}$, the couplings are
\begin{equation}
J_{ij}=
\begin{cases}
J+\lambda/\beta, & \hat p_i\cdot\hat e_{ij}>0,\\
J, & \hat p_i\cdot\hat e_{ij}<0
\end{cases}
\label{eq:MC:interactions}
\end{equation}
where $J>0$ denotes the coupling constant of the reciprocal interactions, $\beta \equiv 1/(k_{\rm B}T)$ defines the inverse temperature ($J=1$ and 
$k_{\rm B} = 1$ hereafter), and $\lambda>0$ is a dimensionless parameter controlling the strength of the NR interactions. The reciprocal Ising model is recovered for $\lambda = 0$. 
\begin{figure}[ht!]
    \centering
    \includegraphics[width=1.0\linewidth]{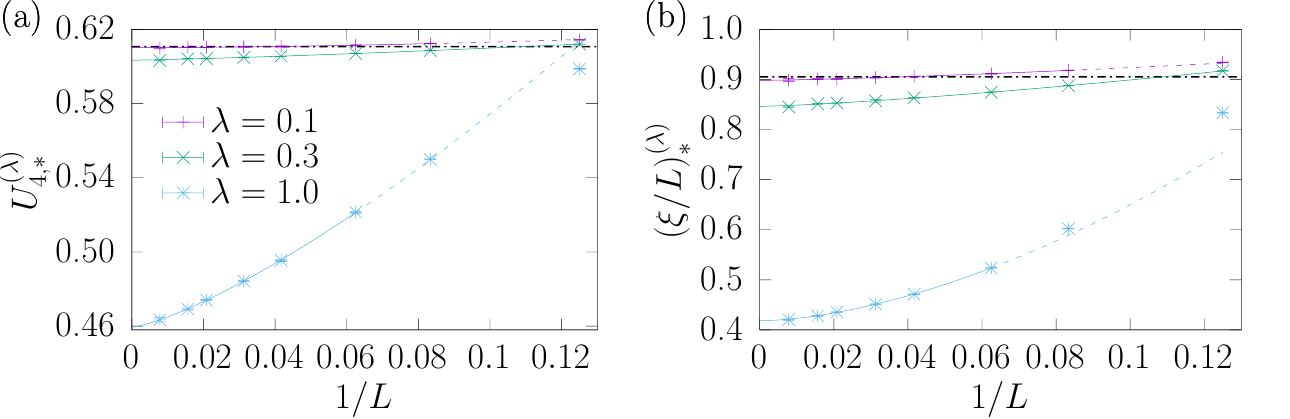}
    \caption{Finite-size scaling extrapolation of the crossing values of (a) the Binder cumulant $U_4$ and (b) the correlation-length ratio $\xi/L$ for pairs of system sizes $(L,2L)$. The horizontal dot-dashed lines indicate the corresponding universal values of the equilibrium $2d$ Ising model.}
    \label{fig:MC:U4_xi}
\end{figure}

We perform MC simulations with single-spin Glauber dynamics~\cite{glauber,barkema_book} on systems with sizes $L \in \{8 - 256\}$ using periodic boundary conditions, considering $\lambda = \{0.1, 0.3, 1\}$. Spins are updated sequentially in typewriter order, with one lattice sweep defining the unit of time. At each update, the local field is computed from the pre-update configuration and the spin flip is accepted with probability $P(\sigma_i\to-\sigma_i) = \frac{1}{2}\left[1-\tanh(\frac{\beta \Delta E_i}{2})\right]$~\cite{Avni25}, where $\Delta E_{i}$ is the change in $E_i$ due to the spin-flip. 

To probe the critical properties of the system, we measure the magnetization $m=N^{-1}\sum_i\sigma_i$, the  susceptibility $\chi=\beta N(\langle m^2\rangle-\langle m\rangle^2)$, and two dimensionless universal quantities: the Binder cumulant $U_4=1-\langle m^4\rangle/(3\langle m^2\rangle^2)$~\cite{binder81} and the second-moment correlation-length ratio $\xi/L$, obtained from the first Fourier mode of the spin field~\cite{amit_book}. For the reciprocal $2d$ Ising universality class on the square lattice with periodic boundary conditions, the corresponding universal values are known with high precision~\cite{salas00}---see also Table~\ref{tab:MC:results}. Although no unique global energy exists, we define the symmetric energy-like observable $e=\frac{1}{2N}\sum_i E_i$, which reduces to the standard Ising energy density in the reciprocal limit and satisfies $\sum_{\{\sigma\}}e(\{\sigma\})=0$, where the sum runs over all spin configurations. We further define the fluctuation quantity $C=\beta^2 N (\langle e^2\rangle-\langle e\rangle^2)$, which should not generally be interpreted through equilibrium fluctuation-dissipation relations. 

We begin by discussing the finite-size scaling results shown in Fig.~\ref{fig:MC:Tc_exponents}, employing hereafter the highly efficient quotients method~\cite{nightingale:76,ballesteros:96,fytas:16b}, which allows for a transparent and controlled analysis of scaling corrections. 
Fig. ~\ref{fig:MC:Tc_exponents}(a) presents the analysis of the crossing temperatures $T_{\ast}$ of pairs $(L,2L)$ of Binder-cumulant curves, using the ansatz $T_\ast^{(\lambda)}=T_{\rm c}^{(\lambda)}+\mathcal{A}_{T}^{(\lambda)}L^{-(1/\nu+\omega)}$, where $T_{\rm c}$ denotes the critical temperature---see Table~\ref{tab:MC:results}---$\mathcal{A}_T$ are nonuniversal scaling amplitudes throughout the present study, $\nu$ is the correlation-length exponent and $\omega$ the leading corrections-to-scaling exponent~\cite{nienhuis:82,blote:88,shao:16,fytas:19}. Our finite-size scaling analysis consistently favors $\omega \approx 4/3$~\cite{nienhuis:82}, which we subsequently fix in the fits shown~\cite{comment_omega}. 
\begin{figure}[ht!]
    \centering
    \includegraphics[width=1.0\linewidth]{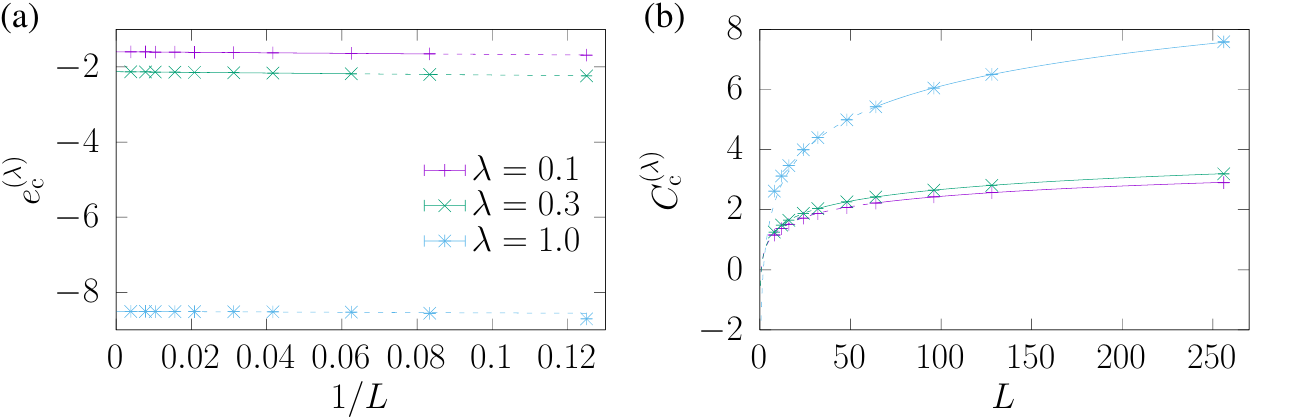}
    \caption{Finite-size scaling behavior of (a) the energy-like observable $e$ and (b) the specific-heat-like fluctuation quantity $C$ at the corresponding critical temperatures. Panel (a) is shown together with simultaneous joint fits for the exponent combination $d-1/\nu$.}
    \label{fig:MC:energy}
\end{figure}
We now turn to the critical exponents: Fig. ~\ref{fig:MC:Tc_exponents}(b) illustrates the effective estimates of the exponent $1/\nu$, obtained from ratios of temperature derivatives of $U_4$ evaluated at the crossings of system sizes $(L,2L)$.  Similarly, panels (c) and (d) display the effective exponents $\beta/\nu$ and $\gamma/\nu$, extracted from the finite-size scaling of the magnetization $m$ and susceptibility $\chi$, respectively. These effective exponents $x = \{1/\nu, \beta/\nu, \gamma/\nu\}$ are expected to approach their asymptotic values according to $x_{\rm eff}^{(\lambda)} = x + \mathcal{A}_{x}^{(\lambda)} L^{-\omega}$~\cite{ballesteros:96,fytas:16b}. Although all three values of $\lambda$ yield critical exponents consistent within error bars, Fig.~\ref{fig:MC:Tc_exponents} clearly demonstrates that the approach to the thermodynamic limit depends strongly on $\lambda$. In particular, the amplitudes and effective rates of the finite-size corrections vary substantially between the different NR regimes. The final estimates of the critical exponents are reported in Table~\ref{tab:MC:results}. 

We next analyze the behavior of the dimensionless quantities $U_4$ and $\xi/L$, which, to the best of our knowledge, have not previously been investigated in detail in the context of NR models. 
Figure~\ref{fig:MC:U4_xi} shows the finite-size behavior of the crossing values of $g = \{U_4, \xi/L\}$ on pairs of system sizes $(L,2L)$. The data are fitted using the scaling form $g_{\ast}^{(\lambda)} = g_\infty^{(\lambda)}+\mathcal{A}_g^{(\lambda)}L^{-\omega}$, where $g_\infty$ denotes the thermodynamic-limit value. For both observables, the extrapolated values differ systematically from those of the reciprocal Ising ferromagnet and display a pronounced dependence on the NR coupling $\lambda$ (see Table~\ref{tab:MC:results}). While the critical exponents remain consistent with the Ising universality class, the dimensionless ratios $U_4$ and $\xi/L$ are known to be more sensitive to microscopic anisotropies and geometrical properties of the system. In the present model, the NR vision-cone interactions introduce an effective directional asymmetry through the dynamically generated local polarization field $\hat p_i$, thereby modifying the structure of correlations at finite scales. Such anisotropic effects can alter finite-size amplitude ratios and scaling functions without destabilizing the underlying equilibrium critical fixed point~\cite{Grinstein85,Bassler94,Henkel01,Selke06,Selke09,Kastening13}. Moreover, the observed dependence on $\lambda$ suggests that the strength of the NR interactions controls the magnitude of these effective anisotropies.
\begin{table}[t]
\caption{Critical temperatures, dimensionless quantities, and energy-like observables for different values of  $\lambda$. The lower panel summarizes the critical exponents obtained from simultaneous joint fits over all values of $\lambda$ considered. 
The numerical estimates are compared with exact~\cite{ferdinand:69,landau_book} and benchmark-quality~\cite{salas00} results of the $2d$ ($\lambda=0$) Ising universality class.}
\label{tab:MC:results}
\begin{ruledtabular}
\begin{tabular}{lcccc}
 & Ising & $\lambda=0.1$ & $\lambda=0.3$ & $\lambda=1$ \\
\hline
$T_{\rm c}$ 
& $2.269185\cdots$
& $2.54799(5)$
& $3.26213(5)$
& $9.154(1)$
\\

$U_{4,\infty}$
& $0.610692\cdots$
& $0.6102(1)$
& $0.6034(1)$
& $0.459(1)$
\\

$(\xi/L)_\infty$
& $0.905048\cdots$
& $0.898(1)$
& $0.8463(4)$
& $0.418(1)$
\\

$e_\infty$
& $-\sqrt{2}$
& $-1.5962(3)$
& $-2.1270(4)$
& $-8.5069(2)$
\\

$C_{\rm reg}$
& $0.138149\cdots$
& $0.16(2)$
& $0.100(3)$
& $-1.01(3)$
\\
\hline

\multicolumn{5}{c}{Critical exponents}
\\
\hline

 & $1/\nu$ & $\beta/\nu$ & $\gamma/\nu$ & $d - 1/\nu$ 
\\

\hline

Joint fits
& $1.000(1)$
& $0.1251(1)$
& $1.7501(7)$
& $1.001(2)$
\\

Ising
& $1$
& $1/8$
& $7/4$
& $1$
\\
\end{tabular}
\end{ruledtabular}
\end{table}

To further probe the non-equilibrium character of the transition, we examine in Fig.~\ref{fig:MC:energy}(a) the finite-size scaling behavior of the energy-like observable at criticality, using the scaling form $e_{\rm c}^{(\lambda)}=e_{\infty}^{(\lambda)}+\mathcal{A}_{e}^{(\lambda)}L^{-(d-1/\nu)}$~\cite{amit_book}. As the NR parameter $\lambda$ increases, the limiting value $e_\infty$ shifts systematically toward lower values, partly due to the increase of $T_{\rm c}$ with  $\lambda$. At the same time, the exponent combination $d - 1/\nu$ remains fully consistent with the $2d$ Ising universality class for all values of $\lambda$ considered---see last column in Table~\ref{tab:MC:results}. Finally, Fig.~\ref{fig:MC:energy}(b) presents the behavior of the specific-heat-like fluctuation quantity. Motivated by the logarithmic divergence of the equilibrium $2d$ Ising specific heat~\cite{ferdinand:69}, we fit the data using $C_{\rm c}^{(\lambda)} = C_{\rm reg}^{(\lambda)}+\mathcal{A}_{C}^{(\lambda)}\log L$. We observe that the regular contribution, $C_{\rm reg}$, decreases significantly with increasing $\lambda$ and eventually becomes negative for sufficiently large $\lambda$. Since $C$ is defined here as a fluctuation observable rather than a thermodynamic response function, the sign of $C_{\rm reg}$ is not constrained by equilibrium stability arguments. Instead, this behavior reflects the strong modification of the non-singular fluctuation background induced by the NR vision-cone interactions. 

Two remarks are in order. (i) While a preliminary numerical study of this model was reported in Ref.~\cite{Garces2025}, the limited system sizes considered there did not allow a reliable determination of the asymptotic scaling regime, leading to inconclusive results regarding a possible $\lambda$ dependence of $\beta/\nu$. In contrast, the present analysis provides compelling evidence that the critical exponents remain consistent with the $2d$ Ising universality class for all $\lambda$ studied. (ii) Fully consistent estimates are obtained independently from quotients analyses at the crossings of the correlation-length ratio $\xi/L$.

\emph{Dynamic RG analysis.---} Following Ref.~\cite{Garces2025}, the lattice model defined in Eq.~(\ref{eq:MC:interactions}) admits a coarse-grained description in terms of a scalar field $\psi(\mathbf{x},t)$
 obeying
\begin{equation}
\partial_t \psi
=
(-r+\mu\nabla^2)\psi
-u\psi^3
+\frac{\lambda}{2}\psi(\,\mathbf{v}\!\cdot\!\nabla)\psi
+\xi,
\label{eq:continuous description}
\end{equation}
where $\xi$ is Gaussian white noise with correlations $\langle
\xi(\mathbf{x},t)\xi(\mathbf{x}',t')
\rangle = 2D\delta^{d}(\mathbf{x}-\mathbf{x}')
\delta(t-t')$.
The first two terms correspond to the standard Ginzburg-Landau $\psi^4$ theory.  
The term $\propto \lambda$ controls the NR  interactions, introducing a self-advection of the field along the direction $\mathbf{v}$, thereby explicitly breaking rotational symmetry. Due to the anisotropic nonlinear term proportional to $\mathbf{v}\!\cdot\!\nabla$, it is convenient to distinguish between longitudinal and transverse elastic constants (with respect to $\mathbf{v}$), denoted by $\mu_{{\parallel}}$ and $\mu_\perp$, respectively. Similar non-equilibrium  terms arise in coarse-grained descriptions of self-organized criticality~\cite{Bak87,HwaKardar89,HwaKardar92}, driven diffusive systems~\cite{Janssen86,BeckerJanssen94}, and Burgers/KPZ-type dynamics~\cite{Medina89}.

Tree-level power counting~\cite{tauber2014critical} shows that both the reciprocal coupling $u$ and the NR coupling $\lambda$ are marginal at the upper critical dimension $d_{\rm c}=4$, becoming relevant perturbations below four dimensions. The stability of the Wilson--Fisher fixed point in the presence of nonreciprocity must therefore be addressed through an explicit RG calculation. 
Following standard methods, in Fourier space~\cite{Garciaojalvo94}, we obtain the one-loop flow equations (see End Matter for details)
\begin{subequations}
\label{eq:RG flow equations}
\begin{align}
    & \frac{dr}{d\ell} = \left(z + K_d \frac{3uD}{r\mu}\Lambda^{d-2}\right)r , \\
    &\frac{d\mu_{\parallel}}{d\ell} = \left[z-2\zeta + K_d \frac{2\lambda^2D}{\mu^3}\Lambda^{d-4}
    \left(\frac{d+1}{d+2}\right)\right]\mu_{\parallel} , \\
    &\frac{d\mu_\perp}{d\ell} = (z-2)\mu_\perp , \\
    &\frac{du}{d\ell} = \left(z-2a-K_d\frac{9uD}{\mu^2}\Lambda^{d-4}\right)u , \\
    &\frac{d\lambda}{d\ell} = \left[z-\zeta-a-K_d\frac{3uD}{\mu^2}\Lambda^{d-4}
    \left(2-\frac{1}{d+2}\right)\right]\lambda , \\
    &\frac{dD}{d\ell} = (2a+z-\zeta-d+1)D ,
\end{align}
\end{subequations}
where $\Lambda$ is the ultraviolet (UV) cut-off,$z$ and $\zeta$ are the dynamic and anisotropy exponents, respectively, $a=(d-2+\eta)/2$, with $\eta$ being the anomalous dimension, and $K_d = \frac{2\pi^{d/2}}{\Gamma(d/2)(2\pi)^d}$.

For $d < 4$, the RG  flow admits only two physically relevant fixed points when $\lambda$ is treated as an external parameter: the Gaussian fixed point and, in $d = 4-\varepsilon$, a Wilson--Fisher-type fixed point given by
\begin{subequations}
\begin{align}
   & z_{\ast} = z_{\ast}^{(0)},\  \zeta_{\ast} = \zeta_{\ast}^{(0)}
    +\kappa_\varepsilon\lambda^2\Lambda^{-\varepsilon}, \label{eq:wf fixed point a}\\
   & a_{\ast} = a_{\ast}^{(0)}
    +\frac12\kappa_\varepsilon\lambda^2\Lambda^{-\varepsilon}, \\
   & u_{\ast} = u_{\ast}^{(0)}
    -\frac{1}{9}\frac{\lambda^2}{\mu}
    \left(\frac{5-\varepsilon}{6-\varepsilon}\right), \\
   & r_{\ast} = r_{\ast}^{(0)}
    +\frac16\kappa_\varepsilon\lambda^2\Lambda^{2-\varepsilon} \label{eq:wf fixed point d},
\end{align}
\end{subequations}
where $z_\ast^{(0)}=2$, $\zeta_\ast^{(0)}=1$, $a_\ast^{(0)}=\frac{1}{2}(d-2)$, and $r_\ast^{(0)}=-\frac{1}{6}(4-d)\mu\Lambda^2$, $u_\ast^{(0)}=(4-d)\frac{\mu^2}{9K_dD}\Lambda^{4-d}$ correspond to the fixed point of the reciprocal theory $(\lambda=0)$, i.e. the Wilson--Fisher fixed point, and $\kappa_{\varepsilon} = \frac{K_{4-\varepsilon }D}{\mu^3}\frac{5-\varepsilon}{6-\varepsilon}$.

\begin{figure}[t]
    \centering
    \includegraphics[width=0.9\linewidth]{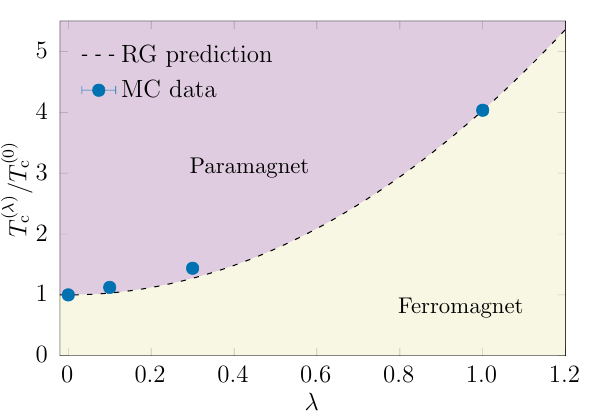}
    \caption{Fit of the critical temperatures obtained from MC simulations and reported in Table~\ref{tab:MC:results}, using the RG prediction $T_{\rm c}^{(\lambda)}/T_{\rm c}^{(0)} - 1 \sim \lambda^{2}$.}
    \label{fig:RG}
\end{figure}

Importantly, in the continuum limit $\Lambda\to\infty$ and for any $d < 4$ ($\varepsilon>0$), the corrections proportional to $\lambda$ vanish, implying $\zeta_{\ast}\to\zeta_{\ast}^{(0)}$ and $a_{\ast}\to a_{\ast}^{(0)}$. The asymptotic critical exponents therefore coincide with those of the reciprocal Wilson--Fisher fixed point despite the explicit violation of reciprocity. The NR perturbation remains RG irrelevant at criticality, in agreement with ~\cite{Bassler94} and fully consistent with our numerical finite-size scaling analysis. In $d=2$, the leading effect of nonreciprocity is instead a shift of the critical mass,
$r_{\ast}-r_{\ast}^{(0)} = \frac16\kappa_2\lambda^2$,
with $\kappa_2=(3/4)K_2D/\mu^3$. Since $r \sim T-T_{\rm c}$~\cite{Garces2025}, the RG analysis predicts the quadratic scaling
$T_{\rm c}^{(\lambda)}-T_{\rm c}^{(0)} \sim \frac16\kappa_2\lambda^2$. As shown in Fig.~\ref{fig:RG}, this prediction is in excellent quantitative agreement with the MC estimates of the critical temperature over the full range of NR couplings studied. Re-expressed in terms of the critical coupling $K_{\rm c} = J / T_{\rm c}$, the RG prediction implies $K_{\rm c} \rightarrow 0$ as $\lambda \rightarrow \infty$, indicating that nonreciprocity alone does not generate spontaneous order but instead shifts the transition to progressively higher temperatures. To the best of our knowledge, this constitutes the first direct quantitative verification, at the microscopic lattice level, of the RG scenario proposed for NR Ising-like systems.

\emph{Conclusions.---} In summary, our combined numerical and RG analysis demonstrates that NR vision-cone interactions do not destabilize the $2d$ Ising fixed point, even though they explicitly break reciprocity and detailed balance. While the critical exponents remain governed by the equilibrium Ising universality class, dimensionless finite-size quantities such as $U_4$ and $\xi/L$ acquire pronounced $\lambda$-dependent corrections, reflecting the directional anisotropy induced by the non-equilibrium interactions. These results indicate that NR perturbations may strongly affect finite-size and geometric properties without modifying the underlying asymptotic universality class. We expect the present work to provide a useful reference point for future studies of criticality and universality in non-reciprocal  systems with directional interactions.

\emph{Acknowledgements.---} Part of the numerical simulations reported in this paper were performed on the High-Performance Computing cluster CERES at the University of Essex. A.~G. thanks Ot Garc\'es and Emir Sezik for helpful discussions and insightful comments. A.~G. acknowledges financial support from AGAUR and the Generalitat de Catalunya under the FI SDUR 2023 program (Ref.~CCI 2021ES05FPR011). M.~C. acknowledges support from the Internationalisation Funds of the TUM Graduate School, which facilitated this collaboration, and partial support from the Fundamental Research Funds for the Central Universities of China (Grant No.~E5ER6601A2). D.~L. acknowledges financial support from MCIU/AEI under Grant No.~PID2022-140407NB-C22. A.~V. and N.G.~F. acknowledge support from the Engineering and Physical Sciences Research Council (Grant No.~EP/X026116/1).  

\bibliography{biblio.bib}

@article{okuma2023non,
   author = "Okuma, Nobuyuki and Sato, Masatoshi",
   title = "Non-Hermitian Topological Phenomena: A Review", 
   journal= "Annu. Rev. Condens. Matter Phys.",
   year = "2023",
   volume = "14",
   number = "Volume 14, 2023",
   pages = "83-107",
   doi = "https://doi.org/10.1146/annurev-conmatphys-040521-033133",
   url = "https://www.annualreviews.org/content/journals/10.1146/annurev-conmatphys-040521-033133",
   publisher = "Annual Reviews",
   issn = "1947-5462",
   type = "Journal Article",
  }

@misc{fruchart2026nonreciprocal,
      title={Nonreciprocal many-body physics}, 
      author={Michel Fruchart and Vincenzo Vitelli},
      year={2026},
      eprint={2602.11111},
      archivePrefix={arXiv},
      primaryClass={cond-mat.stat-mech},
      url={https://arxiv.org/abs/2602.11111}, 
}

@book{landau_book,
  author    = {D. P. Landau and K. Binder},
  title     = {{A Guide to Monte Carlo Simulations in Statistical Physics}},
  edition   = {5},
  year      = {2021},
  publisher = {Cambridge University Press},
  address   = {Cambridge},
  isbn      = {9781108471220}
}

@book{barkema_book,
  author    = {M. E. J. Newman and G. T. Barkema},
  title     = {{Monte Carlo Methods in Statistical Physics}},
  publisher = {Oxford University Press},
  year      = {1999},
  address   = {Oxford},
  isbn      = {9780198517979}
}

@book{amit_book,
  title={{Field theory, the renormalization group, and critical phenomena: graphs to computers}},
  author={Amit, Daniel J and Martín-Mayor, Víctor},
  year={2005},
  publisher={World Scientific Publishing Company}
}

@article{Garces2025,
doi = {10.1088/1742-5468/adc896},
url = {https://doi.org/10.1088/1742-5468/adc896},
year = {2025},
month = {may},
publisher = {IOP Publishing},
volume = {2025},
number = {4},
pages = {043205},
author = {Garcés, Adrià and Levis, Demian},
title = {{Phase transitions in single species Ising models with non-reciprocal couplings}},
journal = { J. Stat. Mech.: Theory Exp.}
}

@book{tauber2014critical,
  title={Critical dynamics: a field theory approach to equilibrium and non-equilibrium scaling behavior},
  author={T{\"a}uber, Uwe C},
  year={2014},
  publisher={Cambridge University Press}
}

@article{nienhuis:82,
  author={Nienhuis, B},
  title={{Analytical calculation of two leading exponents of the dilute Potts model}},
  journal={J. Phys. A.: Math. Gen.},
  volume={15},
  number={1},
  pages={199},
  doi={10.1088/0305-4470/15/1/028},
  url={http://stacks.iop.org/0305-4470/15/i=1/a=028},
  year={1982}
}

@article {shao:16,
	author = {Shao, Hui and Guo, Wenan and Sandvik, Anders W.},
	title = {Quantum criticality with two length scales},
	volume = {352},
	number = {6282},
	pages = {213--216},
	year = {2016},
	doi = {10.1126/science.aad5007},
	publisher = {American Association for the Advancement of Science},
	issn = {0036-8075},
	URL = {http://science.sciencemag.org/content/352/6282/213},
	journal = {Science}
}

@article{blote:88,
  title = {{Corrections to scaling at two-dimensional Ising transitions}},
  author = {Bl\"ote, Henk W. J. and den Nijs, Marcel P. M.},
  journal = {Phys. Rev. B},
  volume = {37},
  issue = {4},
  pages = {1766--1778},
  numpages = {0},
  year = {1988},
  month = {Feb},
  publisher = {American Physical Society},
  doi = {10.1103/PhysRevB.37.1766},
  url = {https://link.aps.org/doi/10.1103/PhysRevB.37.1766}
}

@article{fytas:19,
  title = {{Evidence for Supersymmetry in the Random-Field Ising Model at $D=5$}},
  author = {Fytas, Nikolaos G. and Mart\'{\i}n-Mayor, V\'{\i}ctor and Parisi, Giorgio and Picco, Marco and Sourlas, Nicolas},
  journal = {Phys. Rev. Lett.},
  volume = {122},
  issue = {24},
  pages = {240603},
  numpages = {5},
  year = {2019},
  month = {Jun},
  publisher = {American Physical Society},
  doi = {10.1103/PhysRevLett.122.240603},
  url = {https://link.aps.org/doi/10.1103/PhysRevLett.122.240603}
}

@article{ferdinand:69,
  title = {{Bounded and Inhomogeneous Ising Models. I. Specific-Heat Anomaly of a Finite Lattice}},
  author = {Ferdinand, Arthur E. and Fisher, Michael E.},
  journal = {Phys. Rev.},
  volume = {185},
  issue = {2},
  pages = {832--846},
  numpages = {0},
  year = {1969},
  month = {Sep},
  publisher = {American Physical Society},
  doi = {10.1103/PhysRev.185.832},
  url = {https://link.aps.org/doi/10.1103/PhysRev.185.832}
}

@article{binder81,
  title = {{Critical Properties from Monte Carlo Coarse Graining and Renormalization}},
  author = {Binder, K.},
  journal = {Phys. Rev. Lett.},
  volume = {47},
  issue = {9},
  pages = {693--696},
  numpages = {0},
  year = {1981},
  month = {Aug},
  publisher = {American Physical Society},
  doi = {10.1103/PhysRevLett.47.693},
  url = {https://link.aps.org/doi/10.1103/PhysRevLett.47.693}
}

@article{salas00,
  title={{Universal amplitude ratios in the critical two-dimensional Ising model on a torus}},
  author={Salas, Jes{\'u}s and Sokal, Alan D},
  journal={J. Stat. Phys.},
  volume={98},
  number={3},
  pages={551--588},
  year={2000},
  publisher={Springer},
  doi = {https://doi.org/10.1023/A:1018611122166},
  url = {https://link.springer.com/article/10.1023/A:1018611122166#citeas}
}

@article{Bandini25,
doi = {10.1088/1742-5468/adc243},
url = {https://doi.org/10.1088/1742-5468/adc243},
year = {2025},
month = {may},
publisher = {IOP Publishing},
volume = {2025},
number = {5},
pages = {053205},
author = {Bandini, Gabriele and Venturelli, Davide and Loos, Sarah A M and Jelic, Asja and Gambassi, Andrea},
title = {{The XY model with vision cone: non-reciprocal vs. reciprocal interactions}},
journal = {J. Stat. Mech.: Theory Exp.}
}

@article{Loos23,
  title = {{Long-Range Order and Directional Defect Propagation in the Nonreciprocal $\mathit{XY}$ Model with Vision Cone Interactions}},
  author = {Loos, Sarah A. M. and Klapp, Sabine H. L. and Martynec, Thomas},
  journal = {Phys. Rev. Lett.},
  volume = {130},
  issue = {19},
  pages = {198301},
  numpages = {6},
  year = {2023},
  month = {May},
  publisher = {American Physical Society},
  doi = {10.1103/PhysRevLett.130.198301},
  url = {https://link.aps.org/doi/10.1103/PhysRevLett.130.198301}
}

@article{Rouzaire25,
  title = {{Nonreciprocal Interactions Reshape Topological Defect Annihilation}},
  author = {Rouzaire, Ylann and Pearce, Daniel J. G. and Pagonabarraga, Ignacio and Levis, Demian},
  journal = {Phys. Rev. Lett.},
  volume = {134},
  issue = {16},
  pages = {167101},
  numpages = {6},
  year = {2025},
  month = {Apr},
  publisher = {American Physical Society},
  doi = {10.1103/PhysRevLett.134.167101},
  url = {https://link.aps.org/doi/10.1103/PhysRevLett.134.167101}
}

@misc{rouzaire2026dynamics,
      title={{Dynamics of O(2) excitations in a non-reciprocal medium}}, 
      author={Ylann Rouzaire and Daniel JG Pearce and Ignacio Pagonabarraga and Demian Levis},
      year={2026},
      eprint={2603.23225},
      archivePrefix={arXiv},
      primaryClass={cond-mat.stat-mech},
      url={https://arxiv.org/abs/2603.23225}, 
}

@article{pisegna2024emergent,
author = {Giulia Pisegna  and Suropriya Saha  and Ramin Golestanian },
title = {{Emergent polar order in nonpolar mixtures with nonreciprocal interactions}},
journal = {Proc. Natl. Acad. Sci. U. S. A.},
volume = {121},
number = {51},
pages = {e2407705121},
year = {2024},
doi = {10.1073/pnas.2407705121}
}

@misc{carosi2026time,
      title={Time irreversibility and entropy production in non-Hermitian Model A field theories}, 
      author={Matthias Carosi and Ot Garcés and Adrià Garcés and Demian Levis},
      year={2026},
      eprint={2603.11450},
      archivePrefix={arXiv},
      primaryClass={cond-mat.stat-mech},
      url={https://arxiv.org/abs/2603.11450}, 
}

@article{dadhichi2020nonmutual,
  title = {{Nonmutual torques and the unimportance of motility for long-range order in two-dimensional flocks}},
  author = {Dadhichi, Lokrshi Prawar and Kethapelli, Jitendra and Chajwa, Rahul and Ramaswamy, Sriram and Maitra, Ananyo},
  journal = {Phys. Rev. E},
  volume = {101},
  issue = {5},
  pages = {052601},
  numpages = {13},
  year = {2020},
  month = {May},
  publisher = {American Physical Society},
  doi = {10.1103/PhysRevE.101.052601},
  url = {https://link.aps.org/doi/10.1103/PhysRevE.101.052601}
}

@article{popli2025ordering,
  title = {{Ordering and Defect Cloaking in Nonreciprocal Lattice XY Models}},
  author = {Popli, Pankaj and Maitra, Ananyo and Ramaswamy, Sriram},
  journal = {Phys. Rev. Lett.},
  volume = {135},
  issue = {8},
  pages = {088303},
  numpages = {6},
  year = {2025},
  month = {Aug},
  publisher = {American Physical Society},
  doi = {10.1103/2yky-45sr},
  url = {https://link.aps.org/doi/10.1103/2yky-45sr}
}

@article{Avni25,
  title = {{Nonreciprocal Ising Model}},
  author = {Avni, Yael and Fruchart, Michel and Martin, David and Seara, Daniel and Vitelli, Vincenzo},
  journal = {Phys. Rev. Lett.},
  volume = {134},
  issue = {11},
  pages = {117103},
  numpages = {9},
  year = {2025},
  month = {Mar},
  publisher = {American Physical Society},
  doi = {10.1103/PhysRevLett.134.117103},
  url = {https://link.aps.org/doi/10.1103/PhysRevLett.134.117103}
}

@article{glauber,
    author = {Glauber, Roy J.},
    title = {{Time‐Dependent Statistics of the Ising Model}},
    journal = {J. Math. Phys.},
    volume = {4},
    number = {2},
    pages = {294-307},
    year = {1963},
    month = {02},
    issn = {0022-2488},
    doi = {10.1063/1.1703954},
    url = {https://doi.org/10.1063/1.1703954}
}

@ARTICLE{ballesteros:96,
    AUTHOR = {Ballesteros, H. G. and Fern{\'a}ndez, L. A. and Mart{\'i}n-Mayor, V.
              and Mu{\~n}oz Sudupe, A.},
    TITLE   = {{New universality class in three dimensions?: the antiferromagnetic {RP}$^2$ model}},
    JOURNAL = {Phys. Lett. B},
    SECTION = {B},
    VOLUME  = {378},
    PAGES   = {207},
    YEAR    = {1996},
        DOI     = {10.1016/0370-2693(96)00358-9}
}

@article{nightingale:76,
title = "Scaling theory and finite systems ",
journal = "Physica A: Statistical Mechanics and its Applications ",
volume = "83",
number = "3",
pages = "561 - 572",
year = "1976",
note = "",
issn = "0378-4371",
doi = "http://dx.doi.org/10.1016/0378-4371(75)90021-7",
url = "http://www.sciencedirect.com/science/article/pii/0378437175900217",
author = "M.P Nightingale"
}

@article{fytas:16b,
    author  = {Fytas , Nikolaos G. and Mart\'{\i}n-Mayor, V\'{\i}ctor},
    title   = {{Efficient numerical methods for the random-field Ising model: Finite-size scaling, reweighting extrapolation and computation of response functions}},
    journal = {Phys. Rev. E},
    volume = {93},
    pages = {063308},
    numpages = {15},
    year = {2016},
    month = {June},
    publisher = {American Physical Society},
    doi = {10.1103/PhysRevE.93.063308},
    url = {10.1103/PhysRevE.93.063308}
}

@article{Bak87,
  title = {{Self-organized criticality: An explanation of the 1/f noise}},
  author = {Bak, Per and Tang, Chao and Wiesenfeld, Kurt},
  journal = {Phys. Rev. Lett.},
  volume = {59},
  issue = {4},
  pages = {381--384},
  numpages = {0},
  year = {1987},
  month = {Jul},
  publisher = {American Physical Society},
  doi = {10.1103/PhysRevLett.59.381},
  url = {https://link.aps.org/doi/10.1103/PhysRevLett.59.381}
}

@article{Grinstein85,
  title = {{Statistical Mechanics of Probabilistic Cellular Automata}},
  author = {Grinstein, G. and Jayaprakash, C. and He, Yu},
  journal = {Phys. Rev. Lett.},
  volume = {55},
  issue = {23},
  pages = {2527--2530},
  numpages = {0},
  year = {1985},
  month = {Dec},
  publisher = {American Physical Society},
  doi = {10.1103/PhysRevLett.55.2527},
  url = {https://link.aps.org/doi/10.1103/PhysRevLett.55.2527}
}

@article{Bassler94,
  title = {{Critical Dynamics of Nonconserved Ising-Like Systems}},
  author = {Bassler, K. E. and Schmittmann, B.},
  journal = {Phys. Rev. Lett.},
  volume = {73},
  issue = {25},
  pages = {3343--3346},
  numpages = {0},
  year = {1994},
  month = {Dec},
  publisher = {American Physical Society},
  doi = {10.1103/PhysRevLett.73.3343},
  url = {https://link.aps.org/doi/10.1103/PhysRevLett.73.3343}
}

@article{HwaKardar89,
  title = {{Dissipative transport in open systems: An investigation of self-organized criticality}},
  author = {Hwa, Terence and Kardar, Mehran},
  journal = {Phys. Rev. Lett.},
  volume = {62},
  issue = {16},
  pages = {1813--1816},
  numpages = {0},
  year = {1989},
  month = {Apr},
  publisher = {American Physical Society},
  doi = {10.1103/PhysRevLett.62.1813},
  url = {https://link.aps.org/doi/10.1103/PhysRevLett.62.1813}
}

@article{HwaKardar92,
  title = {{Avalanches, hydrodynamics, and discharge events in models of sandpiles}},
  author = {Hwa, Terence and Kardar, Mehran},
  journal = {Phys. Rev. A},
  volume = {45},
  issue = {10},
  pages = {7002--7023},
  numpages = {0},
  year = {1992},
  month = {May},
  publisher = {American Physical Society},
  doi = {10.1103/PhysRevA.45.7002},
  url = {https://link.aps.org/doi/10.1103/PhysRevA.45.7002}
}

@article{Medina89,
  title = {{Burgers equation with correlated noise: Renormalization-group analysis and applications to directed polymers and interface growth}},
  author = {Medina, Ernesto and Hwa, Terence and Kardar, Mehran and Zhang, Yi-Cheng},
  journal = {Phys. Rev. A},
  volume = {39},
  issue = {6},
  pages = {3053--3075},
  numpages = {0},
  year = {1989},
  month = {Mar},
  publisher = {American Physical Society},
  doi = {10.1103/PhysRevA.39.3053},
  url = {https://link.aps.org/doi/10.1103/PhysRevA.39.3053}
}

@article{BeckerJanssen94,
  title = {{Current-current correlation function in a driven diffusive system with nonconserving noise}},
  author = {Becker, V. and Janssen, H. K.},
  journal = {Phys. Rev. E},
  volume = {50},
  issue = {2},
  pages = {1114--1122},
  numpages = {0},
  year = {1994},
  month = {Aug},
  publisher = {American Physical Society},
  doi = {10.1103/PhysRevE.50.1114},
  url = {https://link.aps.org/doi/10.1103/PhysRevE.50.1114}
}

@article{Janssen86,
    title = {{Field theory of long time behaviour in driven diffusive systems}},
    author = {H. K. Janssen and B. Schmittmann},
    journal = {Z. Phys. B: Condens. Matter},
    volume={63},
    number={4},
    pages={517--520},
     year={1986},
     publisher={Springer},
     doi = {10.1007/BF01726201},
     url = {https://link.springer.com/article/10.1007/BF01726201}
}

@misc{Garciaojalvo94,
      title={{Dynamic Renormalization Group Study of the $\phi^4$ Model with Colored Noise}}, 
      author={J. Garcia-Ojalvo and J. M. Sancho and H. Guo},
      year={1994},
      eprint={cond-mat/9402052},
      archivePrefix={arXiv},
      primaryClass={cond-mat},
      url={https://arxiv.org/abs/cond-mat/9402052}, 
}

@article{Kastening13,
  title = {{Anisotropy and universality in finite-size scaling: Critical Binder cumulant of a two-dimensional Ising model}},
  author = {Kastening, Boris},
  journal = {Phys. Rev. E},
  volume = {87},
  issue = {4},
  pages = {044101},
  numpages = {4},
  year = {2013},
  month = {Apr},
  publisher = {American Physical Society},
  doi = {10.1103/PhysRevE.87.044101},
  url = {https://link.aps.org/doi/10.1103/PhysRevE.87.044101}
}

@article{Selke06,
	author = {Selke, W. },
	date = {2006/05/01},
	doi = {10.1140/epjb/e2006-00209-7},
	id = {Selke2006},
	isbn = {1434-6036},
	journal = {Eur. Phys. J. B},
	number = {2},
	pages = {223--228},
	title = {{Critical Binder cumulant of two-dimensional Ising models}},
	url = {https://doi.org/10.1140/epjb/e2006-00209-7},
	volume = {51},
	year = {2006}}

@article{Henkel01,
doi = {10.1088/0305-4470/34/16/301},
url = {https://doi.org/10.1088/0305-4470/34/16/301},
year = {2001},
month = {apr},
publisher = {},
volume = {34},
number = {16},
pages = {3333},
author = {Malte Henkel and Ulrich Schollwöck},
title = {{Universal finite-size scaling amplitudes in anisotropic
scaling}},
journal = {J. Phys. A: Math. Gen.}
}

@article{Selke09,
  title = {{Critical Binder cumulant in a two-dimensional anisotropic Ising model with competing interactions}},
  author = {Selke, W. and Shchur, L. N.},
  journal = {Phys. Rev. E},
  volume = {80},
  issue = {4},
  pages = {042104},
  numpages = {4},
  year = {2009},
  month = {Oct},
  publisher = {American Physical Society},
  doi = {10.1103/PhysRevE.80.042104},
  url = {https://link.aps.org/doi/10.1103/PhysRevE.80.042104}
}

@article{Klapp23,
	author = {Klapp, Sabine H.  L. },
	date = {2023/01/01},
	doi = {10.1038/s41565-022-01268-0},
	id = {Klapp2023},
	isbn = {1748-3395},
	journal = {Nat. Nanotechnol.},
	number = {1},
	pages = {8--9},
	title = {{Non-reciprocal interaction for living matter}},
	url = {https://doi.org/10.1038/s41565-022-01268-0},
	volume = {18},
	year = {2023}}

@article{Ma25,
  title = {{Rich collective behaviors in nonreciprocal multispecies systems: The interplay between nonreciprocity and permutation symmetry among species}},
  author = {Ma, Weiqiang and Meng, Fei and Cheng, Run and Wang, Jun},
  journal = {Phys. Rev. E},
  volume = {112},
  issue = {2},
  pages = {024210},
  numpages = {12},
  year = {2025},
  month = {Aug},
  publisher = {American Physical Society},
  doi = {10.1103/s6df-sng9},
  url = {https://link.aps.org/doi/10.1103/s6df-sng9}
}

@article{Zia02,
    author = {Zia, R. K. P. and Praestgaard, E. L. and Mouritsen, O. G.},
    title = {{Getting more from pushing less: Negative specific heat and conductivity in nonequilibrium steady states}},
    journal = {Am. J. Phys.},
    volume = {70},
    number = {4},
    pages = {384-392},
    year = {2002},
    month = {04},
    issn = {0002-9505},
    doi = {10.1119/1.1427088},
    url = {https://doi.org/10.1119/1.1427088}
}

@book{tauber_book,
  author    = {Täuber, Uwe C.},
  title     = {{Critical Dynamics: A Field Theory Approach to Equilibrium and Non-Equilibrium Scaling Behavior}},
  publisher = {Cambridge University Press},
  year      = {2014},
  address   = {Cambridge},
  isbn      = {
9781139046213}
}

@incollection{SCHMITTMANN95,
title = {Statistical mechanics of driven diffusive systems},
editor = {B. Schmittmann and R.K.P. Zia},
series = {Phase Transitions and Critical Phenomena},
publisher = {Academic Press},
volume = {17},
pages = {3-214},
year = {1995},
booktitle = {Statistical Mechanics of Driven Diffusive System},
issn = {1062-7901},
doi = {https://doi.org/10.1016/S1062-7901(06)80014-5},
url = {https://www.sciencedirect.com/science/article/pii/S1062790106800145},
author = {B. Schmittmann and R.K.P. Zia}
}

@article{Wilson74,
title = {{The renormalization group and the $\epsilon$ expansion}},
journal = {Phys. Rep.},
volume = {12},
number = {2},
pages = {75-199},
year = {1974},
issn = {0370-1573},
doi = {https://doi.org/10.1016/0370-1573(74)90023-4},
url = {https://www.sciencedirect.com/science/article/pii/0370157374900234},
author = {Kenneth G. Wilson and J. Kogut}
}

@article{Hohenberg77,
  title = {{Theory of dynamic critical phenomena}},
  author = {Hohenberg, P. C. and Halperin, B. I.},
  journal = {Rev. Mod. Phys.},
  volume = {49},
  issue = {3},
  pages = {435--479},
  numpages = {0},
  year = {1977},
  month = {Jul},
  publisher = {American Physical Society},
  doi = {10.1103/RevModPhys.49.435},
  url = {https://link.aps.org/doi/10.1103/RevModPhys.49.435}
}

@book{justin_book,
  author    = {Zinn-Justin, J.},
  title     = {{Quantum Field Theory and Critical Phenomena}},
  publisher = {Oxford University Press},
  year      = {2002},
  address   = {Oxford},
  isbn      = {9780198834625}
}

@MISC{comment_omega,
        TITLE= {{While several numerical studies have reported effective corrections compatible with $\omega=7/4$~\cite{blote:88,shao:16,fytas:19},  our finite-size scaling analysis suggests $\omega \approx 4/3$. This value can be physically motivated from the operator content of the two-dimensional $q$-state Potts conformal field theory. In particular, the operator associated with dilution has scaling dimension $10/3$~\cite{nienhuis:82}, implying a correction exponent $\omega = 10/3 - d = 4/3$, for the present $d = 2$ case.}}
}

\appendix

\section{End Matter} 

\textit{Dynamic RG formulation.}--- We begin by transforming the Langevin equation governing the dynamics of $\psi$ in Eq.~(\ref{eq:continuous description}) to Fourier space. Owing to the anisotropy introduced by the vision--cone interactions, we decompose the elastic constant into two independent components: $\mu_{\parallel}$ along the direction of the vision cone and $\mu_{\perp}$ in the transverse directions.
Using the convention $\psi(\textbf{x},t) = \int \int\frac{d^d \mathbf{k}}{(2\pi)^d}\frac{d\omega}{2\pi}e^{i(\mathbf{k}\cdot \mathbf{x}-\omega t)} \psi(\textbf{k},\omega)$ and defining the bare propagator $G_{\textbf{k},\omega}^{0} = [-i\omega + r+\mu_{\perp}k_{\perp}^2+\mu_{\parallel}k_{\parallel}^2]^{-1}$,
\begin{align}\label{eq:app_a langevin fourier}
    \psi_{\mathbf{q}} = \psi_{\mathbf{q}}^{0} -uG_{\mathbf{q}}^0 \int_{\mathbf{q}_1, \mathbf{q}_2, \mathbf{q}_3}(2\pi)^{d+1}\delta^{d+1}_{\mathbf{q}, \mathbf{q}_1+\mathbf{q}_2+\mathbf{q_3}}\psi_{\mathbf{q}_1}\psi_{\mathbf{q}_2}\psi_{\mathbf{q}_3} \notag \\
    +\frac{1}{2}i\lambda\mathbf{v}\cdot \mathbf{k} G_{
    \mathbf{q}}^0\int_{\mathbf{q}_1, \mathbf{q}_2}(2\pi)^{d+1}\delta^{d+1}_{\mathbf{q}, \mathbf{q}_1+\mathbf{q}_2} \psi_{\mathbf{q}_1}\psi_{\mathbf{q}_2},
\end{align}
\noindent where $\mathbf{q} =(\mathbf{k},\omega)$ and $\int_{\textbf{q}} \equiv \int\frac{d^d\mathbf{k}}{(2\pi)^d}\int_{-\infty}^{\infty}\frac{d\omega}{2\pi}$ and the integral over momentum is performed over $\mathbb{R}^d$. Throughout this Appendix $\omega$ denotes the temporal frequency and should not be confused with the corrections-to-scaling exponent $\omega$ appearing in the finite-size scaling analysis of the main text. For compactness, we use the shorthand notation $\psi_{\mathbf{k},\omega} \equiv \psi_{\mathbf{q}}$, $G_{\mathbf{k},\omega}^0 \equiv G_{\mathbf{q}}^0$ and similarly for all other quantities. Here, $\psi_{\mathbf{q}}^0 = G_{\mathbf{q}}^0\eta_{\mathbf{q}}$ corresponds to the solution of the free theory, while the noise satisfies $\langle \eta_{\mathbf{q}}\rangle =0$, $\langle \eta_{\mathbf{q}} \eta_{\mathbf{q}'}\rangle = 2D(2\pi)^{d+1}\delta^{d+1}(\mathbf{q}+\mathbf{q}')$. It is also convenient to introduce the bare field correlator,
\begin{align}
    &\langle \psi_{\mathbf{q}} \psi_{\mathbf{q}'} \rangle_0 \equiv \langle \psi_{\mathbf{q}}^0\psi_{\mathbf{q}'}^0\rangle = C_{\mathbf{q}}^0 (2\pi)^{d+1}\delta^{d+1}(\mathbf{q}+\mathbf{q}'), \\
    &C_{\mathbf{q}}^0 = 2DG_{\mathbf{q}}^0G_{-\mathbf{q}}^0 = \frac{2D}{\omega^2+(r+\mu_{\perp}k_{\perp}^2+\mu_{\parallel} k_{\parallel}^2)^2} \label{eq:bare correlator}.
\end{align}

The regularization of the theory is naturally provided by the underlying lattice structure of the microscopic model, which introduces a momentum cutoff which strictly depends on the geometrical properties of the lattice. However, since the long--wavelength RG flow is independent of the geometric properties of the lattice, we adopt a spherical cutoff $|\mathbf{k}|\leq \Lambda$, setting $\psi_{\textbf{q}} = 0$ for $|\mathbf{k}|>\Lambda$, for simplicity.
To carry out the dynamic RG analysis, we employ the standard momentum-shell procedure. We first define a reduced cutoff $\Lambda'=e^{-\ell}\Lambda$, with $\ell>0$, and perturbatively solve Eq.~(\ref{eq:app_a langevin fourier}) after integrating out the UV modes in the shell $\Lambda'<|\mathbf{k}|<\Lambda$. The momentum and frequency variables are then rescaled in order to restore the original cutoff, requiring the resulting equation for the infrared (IR) modes to preserve the same form as Eq.~(\ref{eq:app_a langevin fourier}). This procedure defines the scale dependence of the couplings and determines the corresponding RG flow equations. Throughout the perturbative calculation, we adopt a diagrammatic representation.

The vertices associated with the two nonlinear terms in Eq.~(\ref{eq:app_a langevin fourier}) can be represented diagrammatically as follows:

\vspace{-0.3cm}
\begin{equation}\label{eq:feynman rule 4 legged vertex}
\begin{tikzpicture}
  \begin{feynman}
    \vertex (a);
    \vertex [right=1.5cm of a] (b);
    \vertex [right=1cm of b] (c);
    \vertex [above=0.75cm of c] (d);
    \vertex [below=0.75cm of c] (e);
    
    \diagram* {
      (a) -- [fermion, edge label=\(G_\mathbf{q}^0\), very thick] (b),
      (b) -- [double, double distance=1.5pt, edge label=\(\), very thick] (c),
      (b) -- [double, double distance=1.5pt, edge label=\(\), very thick] (d),
      (b) -- [double, double distance=1.5pt, edge label=\(\), very thick] (e),
    };

    \node [below left = 0.1cm and 0.05cm of b] {\( -u\)};
    \node [right=0.1cm of c] {\( \psi_2\)};
    \node [right=0.1cm of d] {\( \psi_1\)};
    \node [right=0.1cm of e] {\( \psi_3\)};
    \node [right=0.7cm of c] {\( = -uG_{\mathbf{q}}^0 \int_{1,2,3} \tilde \delta^{d+1}_{\mathbf{q}, 1+2+3} \psi_1 \psi_2 \psi_3\)};
    \node [circle, fill=blue, inner sep=1.5pt] at (b) {};
    
\end{feynman}
\end{tikzpicture}
\end{equation}
\vspace{-0.5cm}
\begin{equation}\label{eq:feynman rule 3 legged vertex}
\begin{tikzpicture}
  \begin{feynman}
    \vertex (a);
    \vertex [right=1.5cm of a] (b);
    \vertex [right=1.3cm of a] (b');
    \vertex [right=1cm of b] (c);
    \vertex [above=0.75cm of c] (d);
    \vertex [below=0.75cm of c] (e);
    
    \diagram* {
      (a) -- [fermion, edge label=\(G_\mathbf{q}^0\), very thick] (b),
      (b) -- [double, double distance=1.5pt, edge label=\(\), very thick] (d),
      (b) -- [double, double distance=1.5pt, edge label=\(\), very thick] (e),
    };

    \node [below = 0.125cm of b'] {\( \scriptstyle \frac{1}{2}i\lambda (\mathbf{v}\cdot \mathbf{k})\)};
    \node [right=0.1cm of d] {\( \psi_1\)};
    \node [right=0.1cm of e] {\( \psi_2\)};
    \node [right=0.7cm of c] {\( = \frac{1}{2}i\lambda (\mathbf{v}\cdot \mathbf{k})G_{\mathbf{q}}^0 \int_{1,2} \tilde \delta^{d+1}_{\mathbf{q}, 1+2} \psi_1 \psi_2\)};
    \node [circle, fill=blue, inner sep=1.5pt] at (b) {};
    
  \end{feynman}
\end{tikzpicture}
\end{equation}
where the solid line with an arrow and the double lines denote the bare propagator $G_{\mathbf{q}}^0$ and the field $\psi$, respectively. Here we use the shorthand notation $\tilde\delta^{d+1}_{\mathbf{q},\dots} = (2\pi)^{d+1}\delta^{d+1}_{\mathbf{q},\dots}$. The separation between IR and UV modes is implemented by decomposing the field $\psi_{\mathbf{q}}$ according to the momentum scale:
\[
\psi_{\mathbf{q}}=
\begin{cases}
\psi_{\mathbf{q}}^{<}, & |\mathbf{k}|<e^{-\ell}\Lambda,\\
\psi_{\mathbf{q}}^{>}, & |\mathbf{k}|>e^{-\ell}\Lambda.
\end{cases}
\]
That is, $\psi_{\mathbf{q}}^{<}$ contains the IR modes, while $\psi_{\mathbf{q}}^{>}$ contains the UV modes integrated out during the RG procedure. Diagrammatically,
\vspace{-0.2cm}
\begin{equation}
\begin{tikzpicture}
  \begin{feynman}
    \vertex (a1);
    \vertex [right=1.5cm of a1] (a2);
    \vertex [right=1cm of a2] (b1);
    \vertex [right=1.5cm of b1] (b2);
    \vertex [right=1cm of b2] (c1);
    \vertex [right=1.5cm of c1] (c2);
    
    \diagram* {
      (a1) -- [double, double distance=2pt, edge label=\(\psi_{\mathbf{q}}\), very thick] (a2),
      (b1) -- [double, double distance=2pt, ForestGreen, edge label=\(\psi_{\mathbf{q}}^{<}\), very thick] (b2),
      (c1) -- [double, double distance=2pt, BrickRed, edge label={\(\psi_{\mathbf{q}}^{>}\)}, very thick] (c2),
    };

    \node [right=0.2cm of a2] {\(  = \)};
    \node [right=0.2cm of b2] {\(  + \)};
    
  \end{feynman}
\end{tikzpicture}
\end{equation}
and similarly for the propagator and all other fields entering the perturbative expansion. After separating the IR and UV contributions and integrating out the UV modes, the one--loop corrected Langevin equation can be represented diagrammatically as

\begin{widetext}
\vspace{-0.2cm}
\begin{equation}\label{eq:diagrammatic langevin UV integrated}
\begin{tikzpicture}
  \begin{feynman}
    \vertex (a1);
    \vertex [right=1.5cm of a1] (a2);
    \vertex [right=1cm of a2] (b1);
    \vertex [right=1.5cm of b1] (b2);

    \diagram* {
      (a1) -- [double, double distance=1.5pt, ForestGreen, edge label=\(\), very thick] (a2),
      (b1) -- [dashed, ForestGreen, edge label=\(\), very thick] (b2),
    };

    \node [right=0.2cm of a2] {\(  = \)};
    \node [right=0.2cm of b2] {\(  +3 \)};

    \vertex [right=1cm of b2] (i1);
    \vertex [right=1.5cm of i1] (i2);
    \vertex [right=1.5cm of i2] (i3);
    \vertex [above=0.75cm of i2] (loop_top);

    \diagram* {
      (i1) -- [fermion, ForestGreen, very thick] (i2),
      (i2) -- [double, double distance=1.5pt, ForestGreen, very thick] (i3),
      (i2) -- [fermion, BrickRed, bend left=60, very thick] (loop_top),
      (i2) -- [fermion, BrickRed, bend right=60, very thick] (loop_top),
    };
    
    \node [circle, fill=Blue, inner sep=1.5pt] at (i2) {};
    \node [circle, fill=BrickRed, inner sep=1.5pt] at (loop_top) {};
    \node [right=0.2cm of loop_top, BrickRed] {\(  I^{(1)} \)};

    \vertex [right=1cm of i3] (ii1);
    \vertex [right=1cm of ii1] (ii2);
    \vertex [right=1cm of ii2] (ii3);
    \vertex [right=1cm of ii3] (ii4);
    \vertex [right=0.5cm of ii2] (aux);
    \vertex [above=0.75cm of aux] (loop_top_ii);

    \diagram* {
      (ii1) -- [fermion, ForestGreen, very thick] (ii2),
      (ii3) -- [double, double distance=1.5pt, ForestGreen, very thick] (ii4),
      (ii2) -- [fermion, BrickRed, bend left=60, very thick] (loop_top_ii),
      (ii3) -- [fermion, BrickRed, bend right=60, very thick] (loop_top_ii),
      (ii2) -- [fermion, BrickRed, bend right=60, very thick] (ii3),
    };
    
    \node [left=0.25cm of ii1] {\(  +4 \)};
    \node [circle, fill=Blue, inner sep=1.5pt] at (ii2) {};
    \node [circle, fill=Blue, inner sep=1.5pt] at (ii3) {};
    \node [circle, fill=BrickRed, inner sep=1.5pt] at (loop_top_ii) {};
    \node [right=0.3cm of loop_top_ii, BrickRed] {\(  I_{\mathbf{q}}^{(2)} \)};

    \vertex [below=1.5cm of b1] (nr3v1);
    \vertex [right=0.75cm of nr3v1] (nr3v2);
    \vertex [right=0.75cm of nr3v2] (nr3v3);
    \vertex [above=0.75cm of nr3v3] (nr3v4);
    \vertex [below=0.75cm of nr3v3] (nr3v5);

    \diagram* {
      (nr3v1) -- [fermion, ForestGreen, very thick] (nr3v2),
      (nr3v2) -- [double, double distance=1.5pt, ForestGreen, very thick] (nr3v3),
      (nr3v2) -- [double, double distance=1.5pt, ForestGreen, very thick] (nr3v4),
      (nr3v2) -- [double, double distance=1.5pt, ForestGreen, very thick] (nr3v5),
    };

    \node [left=0.1cm of nr3v1] {\( +\)};
    \node [circle, fill=Blue, inner sep=1.5pt] at (nr3v2) {};

    \vertex [below=1.5cm of i1] (iii1);
    \vertex [right=1cm of iii1] (iii2);
    \vertex [right=1cm of iii2] (iii3);
    \vertex [right=1cm of iii3] (iii4);
    \vertex [above=0.75cm of iii3] (iii_aux_1);
    \vertex [above=0.75cm of iii4] (iii_aux_2);
    \vertex [below left = 0.75cm and 0.5cm of iii3] (iii_aux_3);

    \diagram* {
      (iii1) -- [fermion, ForestGreen, very thick] (iii2),
      (iii2) -- [fermion, BrickRed, very thick] (iii3),
      (iii3) -- [double, double distance=1.5pt, ForestGreen, very thick] (iii4),
      (iii2) -- [double, double distance=1.5pt, ForestGreen, very thick] (iii_aux_1),
      (iii3) -- [double, double distance=1.5pt, ForestGreen, very thick] (iii_aux_2),
      (iii2) -- [fermion, BrickRed, very thick, bend right=60] (iii_aux_3),
      (iii3) -- [fermion, BrickRed, very thick, bend left=60] (iii_aux_3),
    };

     \node [left=0.1cm of iii1] {\(  +18 \)};
     \node [circle, fill=Blue, inner sep=1.5pt] at (iii2) {};
     \node [circle, fill=Blue, inner sep=1.5pt] at (iii3) {};
     \node [circle, fill=BrickRed, inner sep=1.5pt] at (iii_aux_3) {};
     \node [right=0.3cm of iii_aux_3, BrickRed] {\(  I^{(3)} \)};

    \vertex [below=2cm of nr3v1] (nr2v1);
    \vertex [right=0.75cm of nr2v1] (nr2v2);
    \vertex [right=0.75cm of nr2v2] (nr2v3);
    \vertex [above=0.75cm of nr2v3] (nr2v4);
    \vertex [below=0.75cm of nr2v3] (nr2v5);

    \diagram* {
      (nr2v1) -- [fermion, ForestGreen, very thick] (nr2v2),
      (nr2v2) -- [double, double distance=1.5pt, ForestGreen, very thick] (nr2v4),
      (nr2v2) -- [double, double distance=1.5pt, ForestGreen, very thick] (nr2v5),
    };

    \node [left=0.1cm of nr2v1] {\( +\)};
    \node [circle, fill=Blue, inner sep=1.5pt] at (nr2v2) {};

    \vertex [below=2cm of iii1] (iv1);
    \vertex [right=1cm of iv1] (iv2);
    \vertex [right=1cm of iv2] (iv3);
    \vertex [right=1cm of iv3] (iv4);
    \vertex [above=0.75cm of iv3] (iv_aux_1);
    \vertex [above=0.75cm of iv4] (iv_aux_2);
    \vertex [above left = 0.75cm and 0.5cm of iv3] (iv_aux_3);
    \vertex [below=0.75cm of iv4] (iv_aux_4);

    \diagram* {
      (iv1) -- [fermion, ForestGreen, very thick] (iv2),
      (iv2) -- [fermion, BrickRed, very thick, bend right = 60] (iv3),
      (iv3) -- [double, double distance=1.5pt, ForestGreen, very thick] (iv_aux_4),
      (iv3) -- [double, double distance=1.5pt, ForestGreen, very thick] (iv_aux_2),
      (iv2) -- [fermion, BrickRed, very thick, bend left=60] (iv_aux_3),
      (iv3) -- [fermion, BrickRed, very thick, bend right=60] (iv_aux_3),
    };

     \node [left=0.1cm of iv1] {\(  +6 \)};
     \node [circle, fill=Blue, inner sep=1.5pt] at (iv2) {};
     \node [circle, fill=Blue, inner sep=1.5pt] at (iv3) {};
     \node [circle, fill=BrickRed, inner sep=1.5pt] at (iv_aux_3) {};

     \node [left=0.3cm of iv_aux_3, BrickRed] {\(  I_{\mathbf{q}}^{(4)} \)};

    \vertex [below=3.5cm of ii1] (v1);
    \vertex [right=1cm of v1] (v2);
    \vertex [right=1cm of v2] (v3);
    \vertex [right=1cm of v3] (v4);
    \vertex [above=0.75cm of v3] (v_aux_1);
    \vertex [above=0.75cm of v4] (v_aux_2);
    \vertex [below left = 0.75cm and 0.5cm of v3] (v_aux_3);

    \diagram* {
      (v1) -- [fermion, ForestGreen, very thick] (v2),
      (v2) -- [fermion, BrickRed, very thick] (v3),
      (v3) -- [double, double distance=1.5pt, ForestGreen, very thick] (v4),
      (v2) -- [double, double distance=1.5pt, ForestGreen, very thick] (v_aux_1),
      (v2) -- [fermion, BrickRed, very thick, bend right=60] (v_aux_3),
      (v3) -- [fermion, BrickRed, very thick, bend left=60] (v_aux_3),
    };

     \node [left=0.1cm of v1] {\(  +12 \)};
     \node [circle, fill=Blue, inner sep=1.5pt] at (v2) {};
     \node [circle, fill=Blue, inner sep=1.5pt] at (v3) {};
     \node [circle, fill=BrickRed, inner sep=1.5pt] at (v_aux_3) {};
     \node [right=0.3cm of v_aux_3, BrickRed] {\(  I_{\mathbf{q}}^{(5)} \)};

    \vertex [below=2cm of nr2v1] (nc11);
    \vertex [right=0.75cm of nc11] (nc12);
    \vertex [right=0.75cm of nc12] (nc13);
    \vertex [right=0.75cm of nc13] (nc14);
    \vertex [above=0.5cm of nc13] (nc1_aux_1);
    \vertex [below=0.5cm of nc13] (nc1_aux_2);
    \vertex [above=0.5cm of nc14] (nc1_aux_3);
    \vertex [below=0.5cm of nc14] (nc1_aux_4);

    \diagram* {
      (nc11) -- [fermion, ForestGreen, very thick] (nc12),
      (nc12) -- [fermion, BrickRed, very thick] (nc13),
      (nc13) -- [double, double distance=1.5pt, ForestGreen, very thick] (nc14),
      (nc13) -- [double, double distance=1.5pt, ForestGreen, very thick] (nc1_aux_3),
      (nc13) -- [double, double distance=1.5pt, ForestGreen, very thick] (nc1_aux_4),
      (nc12) -- [double, double distance=1.5pt, ForestGreen, very thick] (nc1_aux_1),
      (nc12) -- [double, double distance=1.5pt, ForestGreen, very thick] (nc1_aux_2),
    };

     \node [left=0.1cm of nc11] {\(  +3 \)};
     \node [circle, fill=Blue, inner sep=1.5pt] at (nc12) {};
     \node [circle, fill=Blue, inner sep=1.5pt] at (nc13) {};

    \vertex [right=1cm of nc14] (nc21);
    \vertex [right=0.75cm of nc21] (nc22);
    \vertex [right=0.75cm of nc22] (nc23);
    \vertex [right=0.75cm of nc23] (nc24);
    \vertex [above=0.5cm of nc23] (nc2_aux_1);
    \vertex [below=0.5cm of nc23] (nc2_aux_2);
    \vertex [above=0.5cm of nc24] (nc2_aux_3);
    \vertex [below=0.5cm of nc24] (nc2_aux_4);

    \diagram* {
      (nc21) -- [fermion, ForestGreen, very thick] (nc22),
      (nc22) -- [fermion, BrickRed, very thick] (nc23),
      (nc23) -- [double, double distance=1.5pt, ForestGreen, very thick] (nc2_aux_3),
      (nc23) -- [double, double distance=1.5pt, ForestGreen, very thick] (nc2_aux_4),
      (nc22) -- [double, double distance=1.5pt, ForestGreen, very thick] (nc2_aux_1),
      (nc22) -- [double, double distance=1.5pt, ForestGreen, very thick] (nc2_aux_2),
    };

     \node [left=0.1cm of nc21] {\(  +3 \)};
     \node [circle, fill=Blue, inner sep=1.5pt] at (nc22) {};
     \node [circle, fill=Blue, inner sep=1.5pt] at (nc23) {};

    \vertex [right=1cm of nc24] (nc31);
    \vertex [right=0.75cm of nc31] (nc32);
    \vertex [right=0.75cm of nc32] (nc33);
    \vertex [right=0.75cm of nc33] (nc34);
    \vertex [above=0.5cm of nc33] (nc3_aux_1);
    \vertex [below=0.5cm of nc33] (nc3_aux_2);
    \vertex [above=0.5cm of nc34] (nc3_aux_3);
    \vertex [below=0.5cm of nc34] (nc3_aux_4);

    \diagram* {
      (nc31) -- [fermion, ForestGreen, very thick] (nc32),
      (nc32) -- [fermion, BrickRed, very thick] (nc33),
      (nc33) -- [double, double distance=1.5pt, ForestGreen, very thick] (nc3_aux_3),
      (nc33) -- [double, double distance=1.5pt, ForestGreen, very thick] (nc3_aux_4),
      (nc32) -- [double, double distance=1.5pt, ForestGreen, very thick] (nc3_aux_1),
    };

     \node [left=0.1cm of nc31] {\(  +2 \)};
     \node [circle, fill=Blue, inner sep=1.5pt] at (nc32) {};
     \node [circle, fill=Blue, inner sep=1.5pt] at (nc33) {};

    \vertex [right=1cm of nc34] (nc41);
    \vertex [right=0.75cm of nc41] (nc42);
    \vertex [right=0.75cm of nc42] (nc43);
    \vertex [right=0.75cm of nc43] (nc44);
    \vertex [above=0.5cm of nc43] (nc4_aux_1);
    \vertex [below=0.5cm of nc43] (nc4_aux_2);
    \vertex [above=0.5cm of nc44] (nc4_aux_3);
    \vertex [below=0.5cm of nc44] (nc4_aux_4);

    \diagram* {
      (nc41) -- [fermion, ForestGreen, very thick] (nc42),
      (nc42) -- [fermion, BrickRed, very thick] (nc43),
      (nc43) -- [double, double distance=1.5pt, ForestGreen, very thick] (nc44),
      (nc43) -- [double, double distance=1.5pt, ForestGreen, very thick] (nc4_aux_3),
      (nc43) -- [double, double distance=1.5pt, ForestGreen, very thick] (nc4_aux_4),
      (nc42) -- [double, double distance=1.5pt, ForestGreen, very thick] (nc4_aux_1),
    };

     \node [left=0.1cm of nc41] {\(  +2 \)};
     \node [circle, fill=Blue, inner sep=1.5pt] at (nc42) {};
     \node [circle, fill=Blue, inner sep=1.5pt] at (nc43) {};
    
  \end{feynman}
\end{tikzpicture}
\end{equation}

\end{widetext}

The integration of the UV modes is performed by averaging over the bare distribution, such that all the pairs of UV modes are contracted using Wick's theorem and the resulting two--point functions of the contractions correspond to the correlator $C_{\mathbf{q}}^0$ in Eq.~(\ref{eq:bare correlator}). From the diagrammatic representation of the Langevin equation with the UV modes integrated out [Eq.~(\ref{eq:diagrammatic langevin UV integrated})], it becomes clear that, at one--loop order, the renormalization of the reciprocal interaction coupling $u$ is not affected by the presence of the NR interaction. Indeed, the only resulting diagram with three external $\psi^{<}$ legs contains exclusively four--point vertices. Diagrams 1 and 2, associated with the loop integrals $I^{(1)}$ and $I_{\mathbf{q}}^{(2)}$, renormalize the mass and the elastic constants $\mu_{\parallel}$ and $\mu_{\perp}$. Diagram 3, with loop integral $I^{(3)}$, renormalizes the reciprocal coupling $u$, while diagrams 4 and 5, corresponding to the loop integrals $I_{\mathbf{q}}^{(4)}$ and $I_{\mathbf{q}}^{(5)}$, renormalize the NR coupling $\lambda$. All remaining diagrams generate higher--order interaction terms, except for the second one from the end, which would correct the reciprocal interaction coupling. However, since the loop integral contains 3 point vertices, the contribution vanishes when evaluating at $\mathbf{k} = 0$. Note that the noise amplitude $D$ is not renormalized at one-loop order. Its renormalization would follow from the perturbative evaluation of the two-point correlator $\langle \psi_{\mathbf q}\psi_{\mathbf q'}\rangle$. However, a straightforward application of Wick's theorem shows that all one-loop contributions cancel, implying that $D$ receives no correction at this order.

Since the three--point vertex carries momentum, the perturbative expansion generates derivative operators of arbitrary order upon transforming back to real space. Only the contributions of order $0$, $1$ ($\sim \mathbf{k}$), and $2$ ($\sim \mathbf{k}^2$) renormalize the couplings already present in the theory, while higher--order derivative operators are neglected. Choosing $\mathbf{v}=(1,\dots,0)^T$, such that $\mathbf{k}_{\parallel} =(k_{\parallel},\dots,0)^T$, the renormalization of $\lambda$ is determined by the terms linear in $k_x$ arising from diagrams 4 and 5. Similarly, the renormalizations of $\mu_{\perp}$ and $\mu_{\parallel}\equiv\mu_x$ are obtained from the quadratic contributions of diagram 2 in $\mathbf{k}_{\perp}$ and $\mathbf{k}_{\parallel}$, respectively.

After evaluating the loop integrals in Eq.~(\ref{eq:diagrammatic langevin UV integrated}), we rescale the theory in order to restore the original cutoff $\Lambda$. The self-advection term $\propto \lambda$ explicitly breaks rotational invariance, which is reflected in the anisotropic rescaling procedure by singling out the direction selected by the vision cone. Defining $\tilde{\mathbf{k}}_{\perp}=e^{\ell}\mathbf{k}_{\perp}$,
$\tilde{k}_{\parallel}=e^{\zeta\ell}k_{\parallel}$, and 
$\tilde{\omega}=e^{z\ell}\omega$,
one obtains
\[
G_{\mathbf{q}}^0
=
e^{z\ell}G_{\tilde{\mathbf{q}}}^0
=
e^{z\ell}
\left[
-i\tilde{\omega}
+\tilde{r}
+\tilde{\mu}_{\perp}\tilde{k}_{\perp}^2
+\tilde{\mu}_{\parallel}\tilde{q}_{\parallel}^2
\right]^{-1},
\]
with the rescaled parameters
$\tilde{r}=e^{z\ell}r$,
$\tilde{\mu}_{\perp}=e^{(z-2)\ell}\mu_{\perp}$, and
$\tilde{\mu}_{\parallel}=e^{(z-2\zeta)\ell}\mu_{\parallel}$,
where $\zeta$ and $z$ denote the roughness and dynamic critical exponents, respectively. The field rescales as
$\tilde{\psi} =
e^{(a-\zeta-(d-1)-z)\ell}\psi$,
while the couplings transform according to
$\tilde{u}=e^{(z-2a)\ell}u$, $\tilde{\lambda}=e^{(z-\zeta-a)\ell}\lambda$, and
$\tilde{D}=e^{(2a+z-\zeta-(d-1))\ell}D$,
with $a=\frac{1}{2}(d-2+\eta)$,
and $\eta$ the anomalous dimension ($\eta=0$ for $u=0$ and $\lambda=0$). Expanding the exponential factors for small $\ell$, and combining these scaling contributions with the one--loop corrections obtained from Eq.~(\ref{eq:diagrammatic langevin UV integrated}), directly yields the RG flow equations [Eqs.~(\ref{eq:RG flow equations}a)--(\ref{eq:RG flow equations}f)] presented in the main text. In order to perform the loop integrals, we have used that, at leading order in the couplings, $\mu_{\perp}=\mu_{\parallel}=\mu$, since deviations from the tree-level values only appear at higher perturbative orders. We also define
$K_d = \frac{2\pi^{d/2}}{\Gamma(d/2)(2\pi)^d}$.

Imposing the non--renormalization conditions for $D$, $\mu_{\perp}$, and $\mu_{\parallel}$ yields three fixed points. One is nonphysical for $d<4$, as it requires a complex fixed-point value $\lambda^\ast$, although it may become relevant above the upper critical dimension. The remaining fixed points are obtained by treating $\lambda$ as an externally fixed parameter, thereby neglecting its RG flow and focusing on the effect of nonreciprocity on the critical properties of the reciprocal theory. This procedure recovers both the Gaussian fixed point and the Wilson--Fisher-like fixed point [Eqs.~(\ref{eq:wf fixed point a})--(\ref{eq:wf fixed point d})] reported in the main text.

\end{document}